\documentclass[aps,prd,showpacs,eqsecnum,twocolumn,superscriptaddress]
{revtex4-1}
\usepackage{amsmath,amssymb,graphicx,color}
\usepackage{bm}
\usepackage{epsf}

\begin{document}

\title{The dynamical mass ejection from binary neutron star mergers:
  Radiation-hydrodynamics study in general relativity}

\author{Yuichiro Sekiguchi$^1$, 
Kenta Kiuchi$^1$, Koutarou Kyutoku$^2$, and Masaru Shibata}

\affiliation{
Yukawa Institute for Theoretical Physics, 
Kyoto University, Kyoto, 606-8502, Japan \\ 
$^2$Department of Physics,
University of Wisconsin-Milwaukee, P.O. Box 413, Milwaukee,
Wisconsin 53201, USA}

\date{\today}

\begin{abstract}

We perform radiation-hydrodynamics simulations of binary neutron star
mergers in numerical relativity on the Japanese ``K'' supercomputer,
taking into account neutrino cooling and heating by an updated
leakage-plus-transfer scheme for the first time.  Neutron stars are
modeled by three modern finite-temperature equations of state (EOS)
developed by Hempel and his collaborators.  
We find that the properties of the dynamical ejecta of the merger such as total mass, average electron 
fraction, and thermal energy depend strongly on the EOS.
Only for a soft EOS (the so-called SFHo), the ejecta mass exceeds $0.01M_{\odot}$. 
In this case, the distribution of the electron fraction of the ejecta becomes broad due to the shock
heating during the merger. These properties are  well-suited for the production of the solar-like 
$r$-process abundance.
For the other stiff EOS (DD2 and TM1), for which a long-lived massive neutron star is formed
after the merger, the ejecta mass is smaller than $0.01M_{\odot}$, although broad 
electron-fraction distributions are achieved by the positron capture and the neutrino heating.

\end{abstract}


\maketitle


\section{Introduction}

The merger of binary neutron stars (BNS) is
one of the most promising sources of gravitational waves for
advanced LIGO~\cite{LIGO}, advanced VIRGO~\cite{VIRGO}, and KAGRA~\cite{KAGRA}, 
which will start operation in a few years. The recent statistical studies suggest
that these gravitational-wave detectors will observe gravitational
waves from merger events as frequently as $\sim
1$--$100$/yr~\cite{Kalogera,RateLIGO}.  The merger of BNS is also a
promising candidate for the central engine of short-hard gamma-ray
bursts. If gravitational waves are observed simultaneously with them,
a long-standing puzzle on the central engine of short-hard gamma-ray
bursts may be resolved. 

In addition to these aspects, BNS are attracting attentions as the 
nucleosynthesis site of heavy elements by the $r$-process~\cite{Lattimer},
which may proceed in the neutron-rich matter ejected during the merger.
Recent observations of metal-poor stars~\cite{Sneden} strongly suggest 
that there should exist 'main' $r$-stars affected by 'universal' $r$-process 
events in which the resulting abundance is close to that of 
solar-abundance pattern 
for nuclei with the atomic number $Z \gtrsim 38$ ($A \gtrsim 90$).
It has recently been revealed~\cite{Roberts,Wanajo2011} that the supernova 
explosion, which was previously considered to be the most promising
candidate for the site of the $r$-process, may not be a viable origin 
in this regard, and the BNS mergers is getting attention.

Furthermore, a strong electromagnetic emission may accompany the
radioactive decay of the $r$-process elements~\cite{Li:1998bw,KN1,TH} 
and it could be an electromagnetic counterpart of gravitational waves from BNS mergers.
An infrared transient event associated with GRB\,130603B is the first 
candidate for such events~\cite{GRB130603B}.
These facts strongly encourage the community of gravitational-wave
astronomy to explore the $r$-process nucleosynthesis and associated
electromagnetic emission in the BNS merger.

For the quantitative study of these topics, we have to clarify the
merger dynamics, subsequent mass ejection, and physical condition 
of the ejecta, which are necessary to study the nucleosynthesis,
subsequent decay of the heavy elements in the ejecta, and
electromagnetic emission from the ejecta.
For this purpose, we have to perform BNS merger simulations taking into account
both general relativistic gravity and detailed microphysical processes.

For the former, recent numerical relativity simulations (e.g.,~\cite{Hotokezaka:2013iia}; 
see also~\cite{Bauswein} for simulations in approximate general relativistic gravity)
have clarified that the general relativistic gravity can be the key for the mass ejection: 
In general relativity, shock heating plays a prominent role in the 
merger process, and consequently, the ejecta that is dynamically expelled during the merger 
(dynamical ejecta) are composed not only of those driven by the tidal interactions 
but of those driven by the thermal pressure, by contrast with the result in
Newtonian simulations (e.g.,~\cite{Rosswog}) for which the tidal component is major.

For the latter, we have recently developed a neutrino-radiation hydrodynamics code, 
and now, we can perform simulations both employing a wide variety of equations of state (EOS) 
for the nuclear matter in which finite-temperature effects are incorporated and handling 
neutrino cooling and heating with reasonable sophistication.

This is the first study based on these modern aspects of the merger 
dynamics in general relativity taking into account the microphysics.
In this paper, we report the latest result of our simulations 
for equal-mass BNS mergers of typical neutron-star mass ($1.35M_\odot$) 
for three representative EOS, among which the radius of neutron 
stars is appreciably different. 
In this paper, we only consider the case of equal-mass binaries.
The dependence on the mass-ratio and the total mass will be studied in 
a future work.
We will show that the physical properties of the dynamical ejecta such as
the mass and neutron fraction depend strongly on the EOS. 
We find that for producing mildly neutron-rich dynamical ejecta 
of large mass with a broad range of the electron fraction, 
a relatively soft EOS that yields small-radius ($\alt 12$\,km) neutron 
stars is necessary.
Because of such a broad distribution of the electron fraction,
the universal~\cite{Sneden} solar-abundance pattern of 
the $r$-process elements may be reproduced without 
need for the other contributions~\cite{Wanajo}.


\section{Method, EOS, initial models, and grid setup}

\begin{table}[t]
\centering
\begin{tabular}{lccccc}
\hline\hline
EOS  & $n_{0}$ (fm$^{-3}$) & $E_{0}$ (MeV) & $K$ (MeV) & $S$ (MeV) & $L$ (MeV) \\
\hline
SFHo & 0.1583 & 16.19 & 245.5 & 31.57 & 47.10 \\
DD2  & 0.1491 & 16.02 & 242.7 & 31.67 & 55.03 \\
TM1  & 0.145  & 16.3  & 281   & 36.9  & 110.8 \\
\hline\hline
\end{tabular}
\caption{\label{tab1} Characteristic properties of EOS at the nuclear 
saturation density. 
$n_{0}$: the nuclear saturation density.
$E_{0}$: the binding energy.
$K$ : the incompressibility.
$S$ : the symmetry energy.
$L$ : the logarithmic derivative of the symmetry energy.
}
\end{table}

We solve Einstein's equation by the puncture-BSSN (Baumgarte-Shapiro-Shibata-Nakamura) formalism as
before~\cite{BSSN,Sekig}.  The 4th-order finite-differencing scheme is
applied to discretize the field equations.
The radiation
hydrodynamics equations are solved by a recently-developed code which
is updated from the previous version: In this new code, neutrino
transport is computed in a leakage-based scheme~\cite{YS} 
incorporating Thorne's moment formalism with a closure relation 
for a free-streaming component~\cite{Kip}. 
For neutrino heating, absorption on free nucleons is taken into account.

We employ three EOS for nuclear matter derived recently by Hempel and
his collaborators, which are referred to as SFHo~\cite{SFHo}, 
DD2~\cite{DD2}, and TM1~\cite{TM1} in the following. 
TM1 EOS, which is also known as Shen EOS~\cite{Shen}, is based on 
the relativistic mean field theory with a parameter set of Ref.~\cite{Toki}
and have been used widely in both supernova and compact-binary merger simulations.
SFHo EOS is constructed so that the predicted neutron star radius matches 
recent neutron star observations by extending the nonlinear Walecka 
model~\cite{SFHo}. DD2 EOS is based on a relativistic mean field model with 
a density dependent coupling~\cite{Typel}. Some characteristic properties of
EOS are listed in Table~\ref{tab1}.

\begin{table}[t]
\centering
\begin{tabular}{lccccc}
\hline\hline
Model & $R_{1.35}$\,(km) & $\Delta x_{9}$\,(m) & $N$ & $M_{\rm ej}\,(M_\odot)$ & $\langle Y_{e} \rangle$\\
\hline
SFHo (high)    &11.9& 150  & 285  & $1.1 \times 10^{-2}$ & 0.31 \\
SFHo (low)     &    & 250  & 160  & $1.3 \times 10^{-2}$ & 0.32 \\
SFHo (no-heat) &    & 250  & 160  & $1.0 \times 10^{-2}$ & 0.29 \\ \hline
DD2  (high)    &13.2& 160  & 285  & $2.1 \times 10^{-3}$ & 0.29 \\
DD2  (low)     &    & 270  & 160  & $1.9 \times 10^{-3}$ & 0.29 \\
DD2  (no-heat) &    & 270  & 160  & $0.9 \times 10^{-3}$ & 0.26 \\ \hline
TM1  (high)    &14.5& 200  & 285  & $1.2 \times 10^{-3}$ & 0.26 \\ 
TM1  (low)     &    & 300  & 160  & $0.8 \times 10^{-3}$ & 0.25 \\
\hline\hline
\end{tabular}
\caption{\label{tab2} $R_{1.35}$: the radius of spherical neutron
  stars of mass $1.35M_\odot$.  $\Delta x_{9}$: the grid spacing in
  the finest refinement level. $N$: the grid number in one positive
  direction for each refinement level.  $M_{\rm ej}$ and 
  $\langle Y_{e} \rangle$ denote the ejecta mass and the averaged value 
  of $Y_{e}$ measured at the end of the simulations.
  Model name follows the EOS. }
\end{table}

For all of them, the predicted maximum mass
for spherical neutron stars is larger than the largest well-measured
mass of neutron stars, $\approx 2M_\odot$~\cite{Demorest:2010bx}.  For
these EOS, the radius of neutron stars with mass $1.35M_\odot$ is
$R_{1.35}=11.9$~km (SFHo), $13.2$~km (DD2), $14.5$~km (TM1),
respectively (see Table~\ref{tab2}). 
We refer to an EOS with a small neutron star radius ($R_{1.35} \leq 12$~km) 
like SFHo as a {\it soft} EOS and an EOS with a large radius ($R_{1.35} \gtrsim 13$~km) 
as a {\it stiff} EOS.
The stellar radius plays a key role
for determining the merger remnant and the properties of the dynamical ejecta.

In numerical simulations, we have to follow the ejecta with velocity
0.1--$0.3c$ ($c$ is the speed of light), which expand to $> 10^3$\,km
in the simulation time. To follow the ejecta motion as well as to
resolve neutron stars, we employ a fixed mesh-refinement algorithm. In
this work, we prepare 9 refinement levels with the varying grid
spacing as $\Delta x_l=2^{9-l}\Delta x_9$ ($l=1, 2, \cdots, 9$) and
all the refinement levels have the same coordinate origin. Here,
$\Delta x_l$ is the grid spacing for the $l$-th level in the Cartesian
coordinates.  For each level, the computational domain covers the
region $[-N \Delta x_{l}, N\Delta x_{l}]$ for $x$- and $y$-directions,
and $[0, N\Delta x_{l}]$ for $z$-direction (the reflection symmetry
with respect to $z=0$ is imposed).  In the highest-resolution run, we
assign $N=285$, $\Delta x_9=150$--200\,m, and utilize $\approx
7,000$~CPUs on the K computer.

To check that the numerical results depend only weakly on the grid 
resolution, we also performed lower-resolution simulations.  
For this case, $N=160$ and $\Delta x_9=250$--300\,m.
As listed in Table~\ref{tab2}, we found that the results such 
as total ejecta mass and averaged values of $Y_{e}$ depend 
very weakly on the grid resolution.
Furthermore, to confirm the importance of the neutrino heating, we also
performed simulations in which the neutrino absorption is switched 
off (denoted as 'no-heat' in Table~\ref{tab2}) and compared the results 
for the first time.

\begin{figure}[t]
\epsfxsize=3.in
\leavevmode
\epsffile{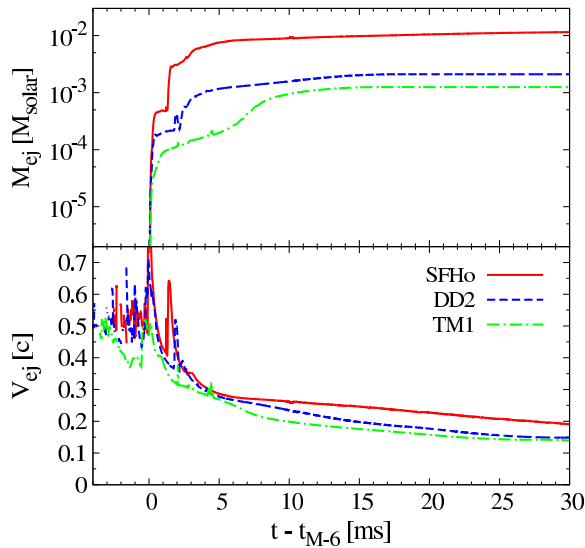}
\caption{Mass (upper panel) and characteristic velocity (lower panel) of the ejecta as
functions of time for SFHo (red solid), DD2 (blue dashed), and TM1 (green dotted-dashed). 
$t_{M-6}$ approximately denotes the time at the onset of merger (see text).
\label{fig1}}
\end{figure}

We consider equal-mass BNS with each mass $1.35M_\odot$.  Observed
neutron stars in BNS typically have the mass ratio close to unity and
the mass in the range $1.20$--$1.45M_\odot$~\cite{Lorimer}. Thus, our
choice reasonably reflects the observational fact. The initial orbital
separation is chosen so that the orbital angular velocity, $\Omega$,
satisfies $Gm_0\Omega/c^3=0.028$ where $m_0=2.7M_\odot$ is the sum of
each mass in isolation and $G$ gravitational constant, respectively.
Table~\ref{tab2} lists the key parameters of our models and simulation
setup.

\section{Results}

\begin{figure*}[t]
\begin{center}
    \includegraphics[scale=0.59]{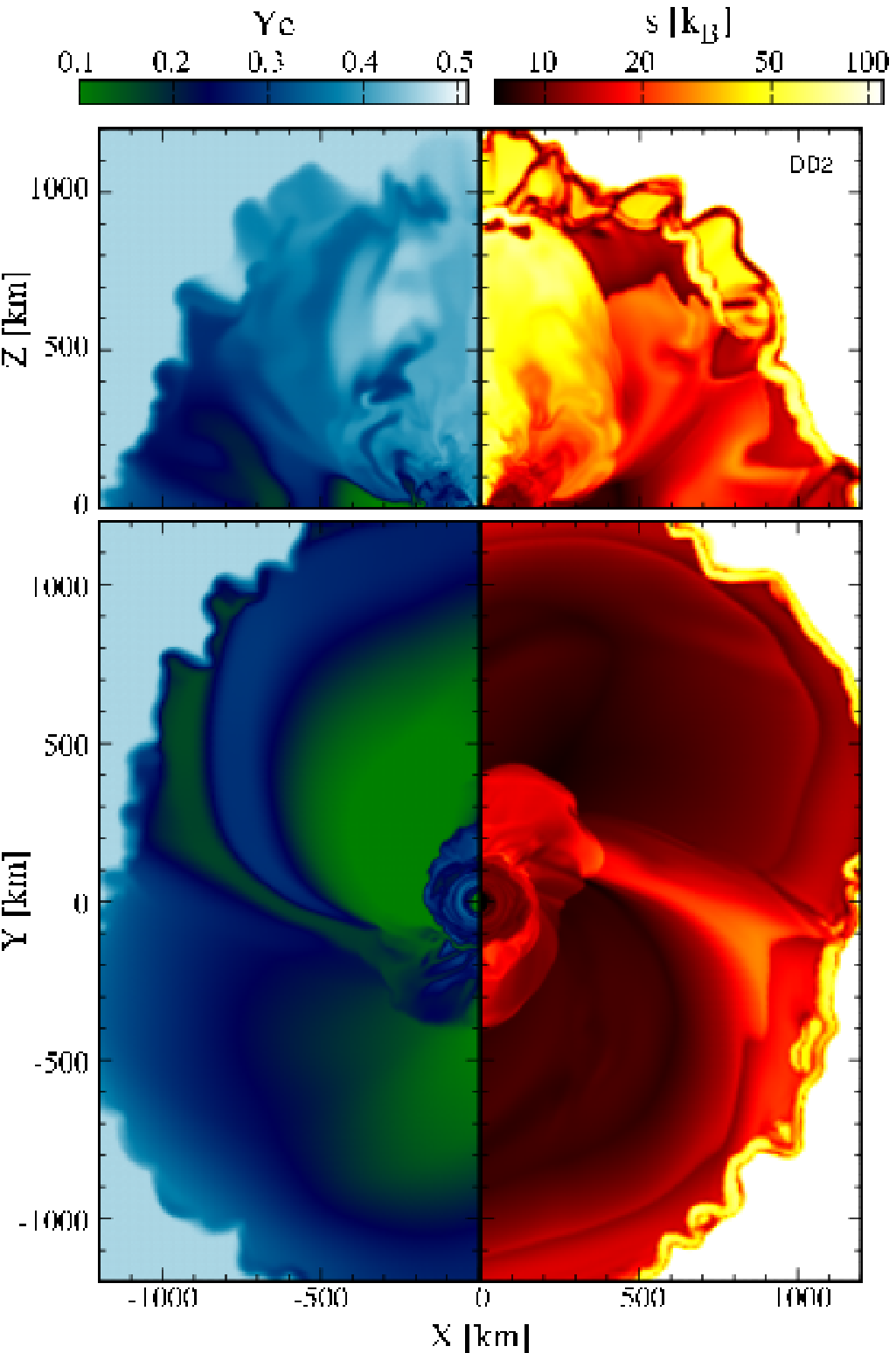}
    \includegraphics[scale=0.59]{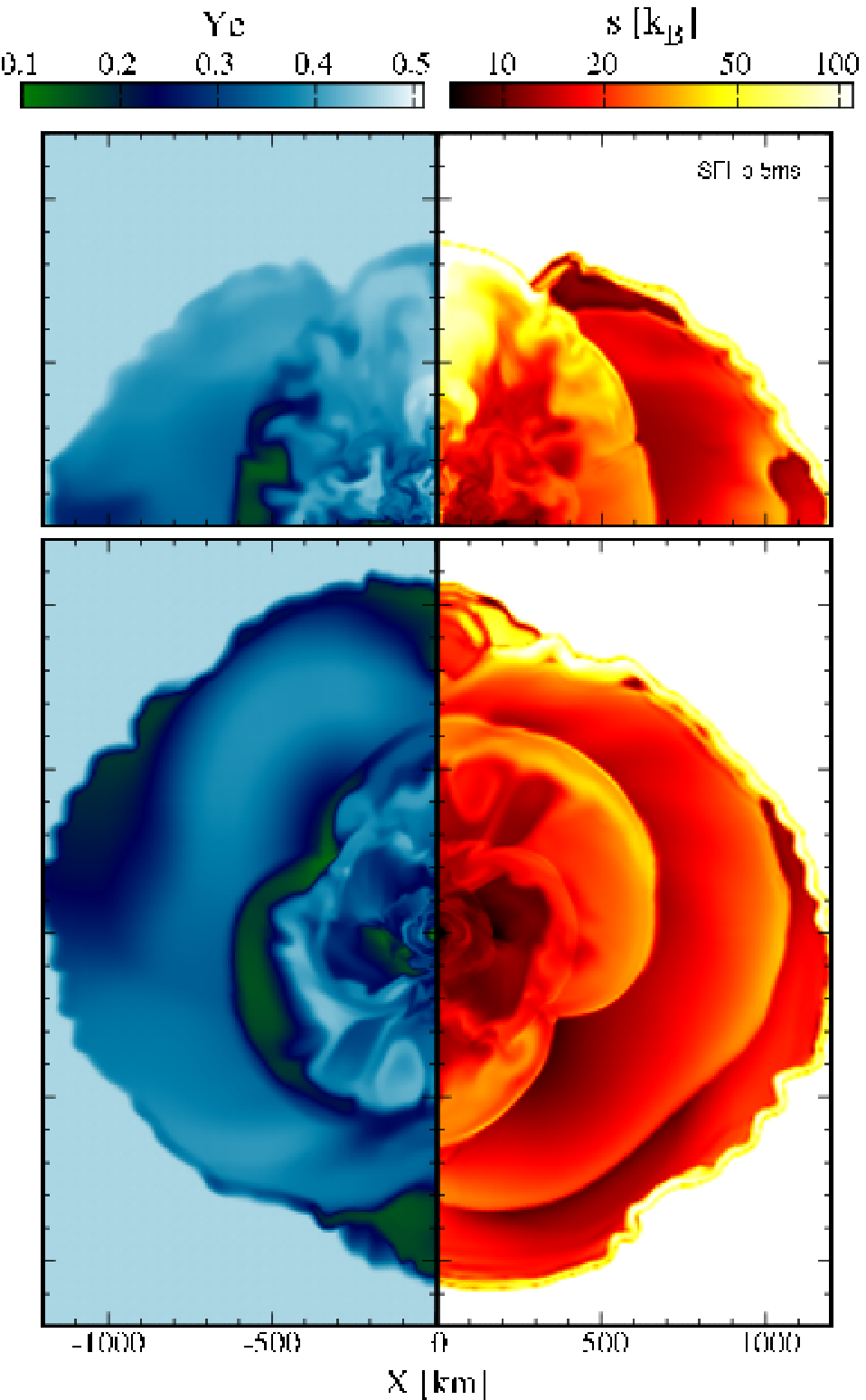}
    \includegraphics[scale=0.59]{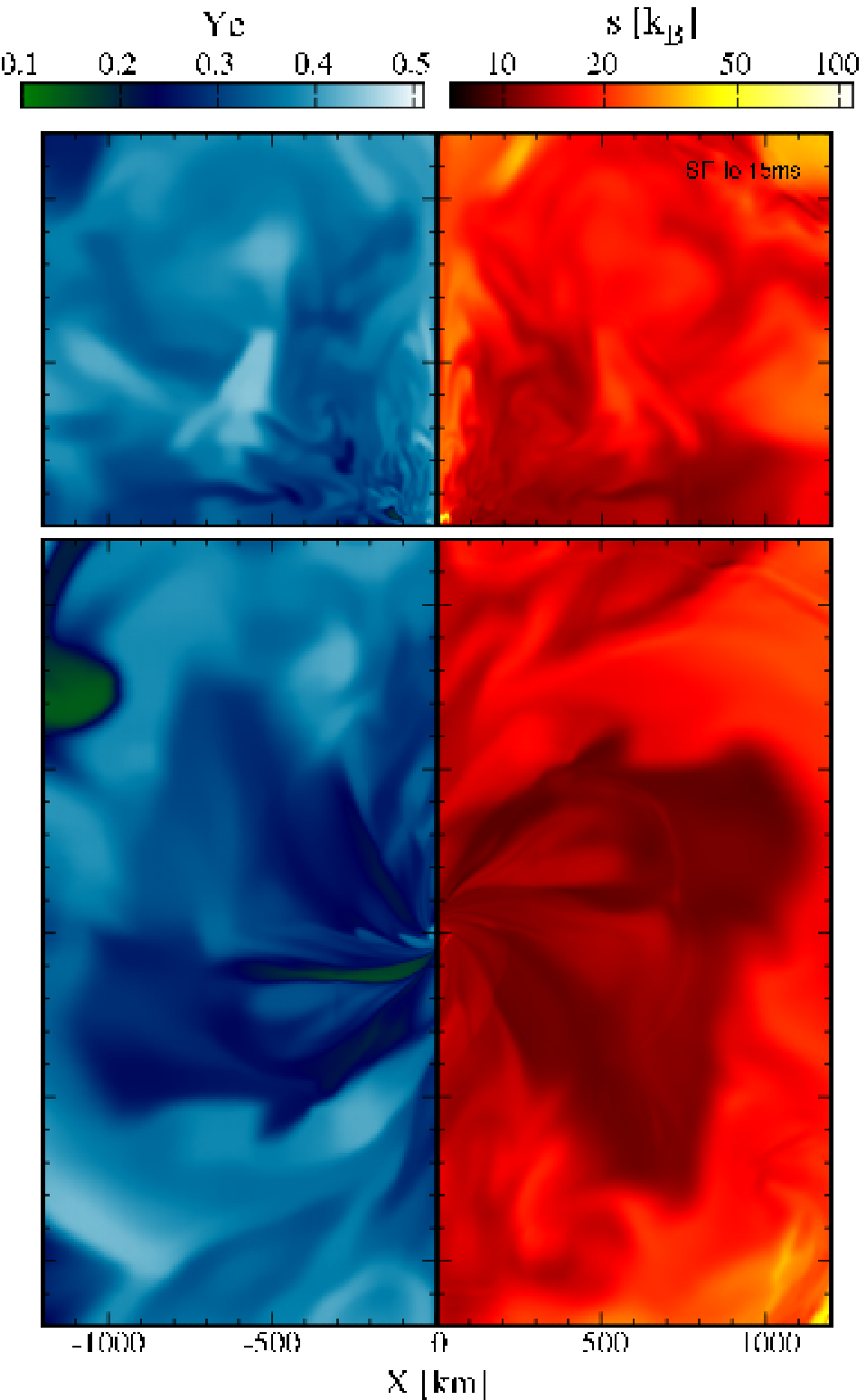}
\end{center}
\vspace{-3mm}
\caption{Contours of the electron fraction, $Y_{e}$, (left half)
  and the entropy per baryon, $s$, (right half) in $x$-$y$ (lower) and
  $x$-$z$ (upper) planes.  {\it left panel}: for DD2 at 8.5\,ms after
  the merger.  {\it middle panel}: for SFHo at 5.0\,ms after the
  merger.  {\it right panel}: for SFHo at 15.0\,ms after the
  merger. 
\label{fig2}}
\end{figure*}

For all the models, a massive neutron star (MNS) is
formed after the onset of merger as expected from our previous
results~\cite{Hotoke2013b}.  The MNS are long-lived in the sense that
their lifetime is much longer than their rotation period of $\alt 1$\,ms.
For SFHo, the MNS eventually collapses to a black hole (BH) in $\sim 10$\,ms 
because the maximum mass of spherical neutron stars is relatively small as 
$\approx 2.0 M_{\odot}$.
The mass and spin parameter of the BH are
$M_{\rm BH} \approx 2.6 M_{\odot}$ and $a_{\rm BH} \approx 0.70$, and a torus with
mass $M_{\rm torus} \approx 0.05M_{\odot}$ is formed around it.
Such a system may be a central engine of short-hard gamma-ray bursts.
For other two cases, the remnant MNS does not collapse to
a BH in our simulation time $\sim 30$--40\,ms. Because the maximum
mass of spherical neutron stars for DD2 and TM1 is $\approx 2.4$ and
$2.2M_\odot$, the formed hot and rapidly rotating MNS with mass $\sim 2.6M_\odot$ 
will not collapse to a BH unless a substantial
fraction of the angular momentum and thermal energy is dissipated by
some transport process and the neutrino emission, respectively 
(e.g.,~\cite{Sekig,Hotoke2013b}).

Figure~\ref{fig1} plots the evolution of the rest mass $M_{\rm ej}$ and the
characteristic velocity $V_{\rm ej}$ for the ejecta.  Here, $t_{M-6}$
denotes the time at which $M_{\rm ej}$ exceeds $10^{-6}M_{\odot}$ 
(hereafter we will use $t_{M-6}$ as the time at the onset of merger). 
We specify the matter as the ejecta if the time component of the fluid
four velocity $u_t$ is smaller than $-1$. Note that another 
condition~\cite{Narayan} for the ejecta $hu_{t} < -1$ where $h$ is the 
specific enthalpy, which may be more appropriate for the hot matter, 
gives slightly larger ejecta mass.
$V_{\rm ej}$ is defined by $\sqrt{2E_{\rm kin}/M_{\rm ej}}$ where $E_{\rm kin}$ 
is kinetic energy of the ejecta. 
Figure~\ref{fig1} shows that the ejecta mass depends strongly
on the EOS: For softer EOS (i.e., for smaller values of $R_{1.35}$),
the ejecta mass is larger. Remarkable is that with the decrease of
$R_{1.35}$ by $\sim 3$\,km, the ejecta mass increases by more than one
order of magnitude and only for $R_{1.35}\alt 12$\,km the ejecta mass
exceeds $0.01M_\odot$, as already indicated in~\cite{Bauswein,Hotokezaka:2013iia}.
The averaged ejecta velocity is $\sim 0.1$--$0.2c$ as also found in~\cite{Bauswein,Hotokezaka:2013iia}.
In the later phase, the total ejecta mass relaxes approximately to a
constant, and the ejecta are in a free expansion phase for all
the models.

There are two major mass ejection mechanisms during the merger phase. 
One is tidal interaction and the other is shock heating. By the tidal
interaction, the matter tends to be ejected near the orbital plane. On
the other hand, by the shock heating, the matter is ejected in a
quasi-spherical manner.  Because both effects play a role, the ejecta
usually have a spheroidal morphology.  For small values of $R_{1.35}$,
the shock heating plays a stronger role and the ejecta in
this case have a quasi-spherical morphology.

\begin{figure}[t]
\epsfxsize=3.in
\leavevmode
\epsffile{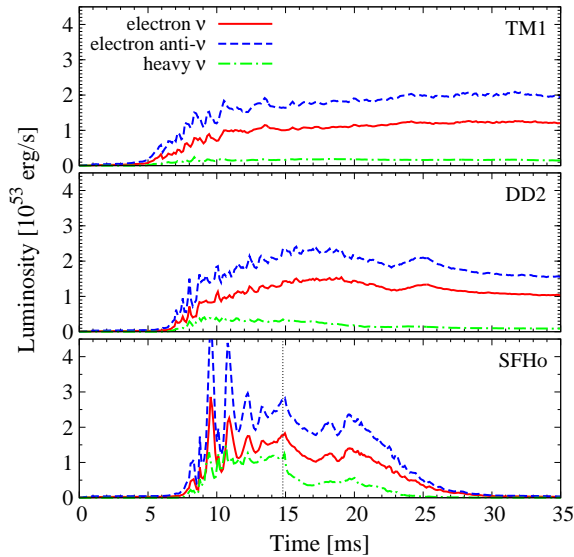}
\caption{Luminosity curves of $\nu_{e}$ (red solid), $\bar{\nu}_{e}$
  (blue dashed), and heavy (green dotted-dashed) neutrinos for TM1
  (top), DD2 (middle), and SFHo (bottom).
\label{fig3}}
\end{figure}

Figure~\ref{fig2} plots the profiles of the electron fraction,
$Y_e$, (left half) and entropy per baryon, $s$, (right half) of the
ejecta on the $x$-$y$ and $x$-$z$ planes for DD2 (left panel) and SFHo
(middle and right panels).  For DD2, the ejecta are composed of (i)
tidally-ejected matter with low values of $Y_{e}$ and $s$ near the
orbital plane and (ii) shock-heated matter with relatively high values
of $Y_{e}$. The shock-heated ejecta are less neutron-rich because 
the temperature gets much higher than $\sim 1$\,MeV as a
result of the shock heating, producing copious $e^-e^+$ pairs that 
activate $e^-$ and $e^+$ captures by protons and neutrons, respectively.  
As a result of $e^-$ and $e^+$ captures, the luminosities of $\nu_{e}$ 
and $\bar{\nu}_{e}$ become quite high as $\gtrsim 10^{53}$\,ergs/s (see Fig.~\ref{fig3}),
as long as the remnant MNS is present. 
Because the original ejecta are neutron-rich, $e^+$ capture dominates 
$e^-$ capture, and hence, the luminosity of $\bar{\nu}_{e}$ is 
higher than that of $\nu_{e}$~\cite{Sekig} and the ejecta become less neutron-rich.

In addition to the tidal-driven and shock-heated components explained above, 
we found the third component in a later phase, that is, {\em neutrino-heated} 
component with even higher values of $Y_{e}$ and $s$ in the region above 
the MNS pole (see the high-entropy region in the left panel ($x$-$z$ plot) 
of Fig.~\ref{fig2}). Furthermore, some fraction of the material obtains 
enough energy to be additional {\it neutrino-driven} ejecta.
Possible existence of such a component was recently reported 
in a MNS system~\cite{Dessart,Perego2014} and a BH and torus system which is expected to 
be formed after the BNS mergers~\cite{Just2014}. 
We confirmed the existence of the neutrino-driven component in self-consistent
numerical-relativity simulations of the merger for the first time.

For TM1, the results are basically similar to those for DD2 except for
the fact that the tidally-ejected component is more dominant and 
the $e^{+}$ capture is less efficient.
Also, the neutrino-driven wind appears to play a major role for the 
mass ejection (see the curve for $t-t_{M-6} > 5$\,ms of Fig.~\ref{fig1})
because the total ejecta mass for this EOS is rather small.
Here, note that it is not easy to exclude the effect of artificial 
atmosphere in grid-based simulations, in particular when the ejecta mass is 
low ($\lesssim 10^{-3} M_{\odot}$) as in the case of TM1. 
The contamination in mass would be $\sim 10^{-4} M_{\odot}$ when the ejecta expand 
to $\sim 2000$\,km in our setting of the atmosphere with density $\sim 10^{3}$\,g/cm$^{3}$, 
while it would be of order of percent if the ejecta is as massive as $\sim 10^{-2}M_{\odot}$. 
The contamination in $Y_{e}$ would be similar level. 
For this reason, in the following, we will basically consider DD2 as a representative of 
a stiff (or moderately stiff) EOS.

For SFHo, shock waves are formed for several times during the merger phase 
as the MNS oscillates with a high amplitude, and hence, a certain fraction of matter 
originally ejected by the tidal interaction is subsequently heated up 
by shocks ($s$ increases), resulting in the increase of the values 
of $Y_e$ via weak interactions. 
On the other hand, other parts less influenced by the shock heating
preserve the neutron-rich nature of the original neutron stars.
As a result of these two facts, the ejecta can have higher 
values of $s$ and $Y_e$ than for DD2 and TM1 even in the orbital plane 
with an appreciably inhomogeneous distribution of $Y_e$
(see the middle panel of Fig.~\ref{fig2}).
Because a BH is formed at $\sim 10$\,ms after the onset of merger 
for SFHo, the strong neutrino emission region is swallowed into 
the BH and neutrino luminosity decreases to 
$\lesssim 10^{53}$\,ergs/s. Hence, there is less clear neutrino-driven
ejecta component for this EOS (see the bottom panel of Fig.~\ref{fig3}).

\begin{figure}[t]
\epsfxsize=3.in
\leavevmode
\epsffile{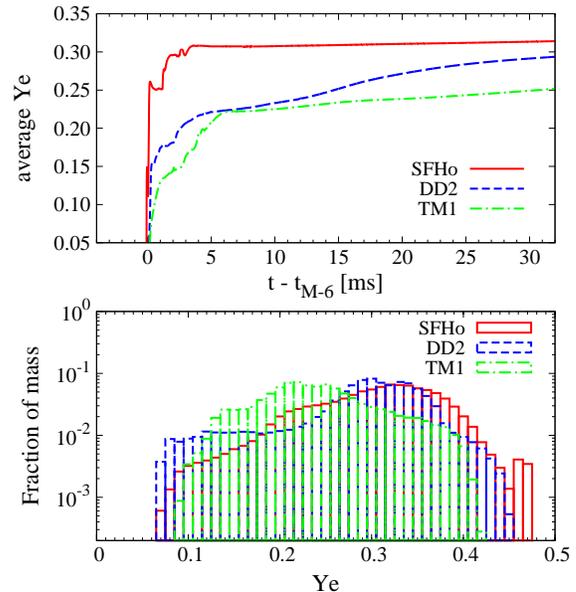}
\caption{ 
{\it Upper panel}: The time evolution of the averaged value of $Y_{e}$
for SFHo (red solid), DD2 (blue dashed), and TM1 (green dotted-dashed).
{\it Lower panel}: The mass-distribution histograms of $Y_e$ normalized by the
total mass of ejecta measured at $\approx 25$\,ms after the onset of merger
for SFHo, DD2, and TM1.
\label{fig4}}
\end{figure}

The upper panel of Fig.~\ref{fig4} shows the time evolution of averaged 
values of $Y_{e}$ ($\langle Y_{e} \rangle$) from which the effect on $Y_{e}$ of the 
shock heating and the resulting positron capture can be seen more clearly.
The several distinct changes in $\langle Y_{e} \rangle$ observed for SFHo in 
$\lesssim 5$\,ms after the onset of merger reflect the strong $e^{+}$ capture
activated by the shock heating. 
During this phase, $\langle Y_{e} \rangle$ for SFHo increases drastically 
to be $\approx 0.3$. 
After this phase, on the other hand, $\langle Y_{e} \rangle$ for SFHo is 
approximately constant because the $e^{-}$ and $e^{+}$ captures balances and 
because the neutrino luminosity decreases to be $\sim 10^{52}$\,ergs/s due to 
the BH formation, which is not sufficient to change $\langle Y_{e} \rangle$
of the massive ejecta.
Thus, for softer EOS like SFHo, $Y_{e}$ is likely to be increased primarily by 
the $e^{+}$ capture.

On the other hand, $\langle Y_{e} \rangle$ for DD2 and TM1 in the early stage 
is low as $Y_{e} \lesssim$ 0.1--0.2, while it increases in time. This is simply because
the shock heating at the first contact is not strong enough to increase $\langle Y_{e} \rangle$ 
significantly for these stiffer EOS; i.e, the original composition of the ejecta 
driven by tidal torque, which is composed primarily of neutron-rich matter with low 
temperature, is temporally preserved as found in~\cite{Rosswog,Bauswein}. 
In the later phase, however, the ejecta become less neutron-rich. 
This is partly due to the positron capture discussed above. 
In addition, the electron neutrinos emitted from the remnant MNS convert some fraction 
of neutrons to protons via the electron neutrino capture (see below for a more detailed
discussion).
For stiffer EOS, the importance of the electron neutrino capture in
increasing $Y_{e}$ of the ejecta is enhanced because of their lower temperature and the 
maintained high neutrino luminosity from the long-lived MNS.

The lower panel of Fig.~\ref{fig4} plots
the mass-distribution histograms for $Y_e$ normalized by the total mass 
of the ejecta at $\approx 25$\,ms after the onset of merger.
For all of the models, $Y_{e}$ is distributed in a broad range 
between $\sim 0.05$ and $0.45$. This result is completely different from that
found in the previous studies~\cite{Rosswog,Bauswein} in which the distribution of 
$Y_{e}$ is very narrow with a lower average value $\lesssim 0.1$. This disparity can be
explained as follows.

In the previous approximate general relativistic study~\cite{Bauswein}, the weak
interaction processes were not taken into account, and hence, the ejecta remain neutron-rich
because there is no way to change $Y_{e}$. In the previous Newtonian studies~\cite{Rosswog}, 
they took into account the neutrino cooling ($e^{-}$ and $e^{+}$ captures). However, as we mentioned already,
the effect of the shock heating is underestimated significantly in Newtonian gravity, 
and hence, the effect of the $e^{+}$ capture would be much weaker than that in our simulations due
to the underestimated temperature.
In addition, they did not take into account the neutrino heating (absorptions) which is expected
to play a role for stiffer EOS in which the positron capture is relatively less important due to 
lower temperature.

To see the effects of the neutrino heating more quantitatively, we
performed simulations without (no-heat) neutrino heating for SFHo and DD2.
We found that for both EOS, the contribution of the neutrino-driven component 
in the ejecta mass is $\sim 10^{-3} M_{\odot}$ at the end of the simulation
(see Table~\ref{tab2}), which is consistent with that found in~\cite{Perego2014}.
The amount of the neutrino-driven ejecta is minor for SFHo
but comparable to the amount of the dynamical ejecta for DD2.
This result suggests that the neutrino heating plays a relatively more important 
role for stiffer EOS like DD2 and TM1 in which the amount of the dynamical ejecta 
is $\sim 10^{-3} M_{\odot}$.

The neutrino heating plays an important role in changing the chemical 
composition ($Y_{e}$) of the ejecta.
As shown in Fig.~\ref{fig3}, the luminosities of $\nu_{e}$
and $\bar{\nu}_{e}$ are quite high as $\gtrsim 10^{53}$\,ergs/s. 
Due to the absorption of neutrinos with this high luminosity, the ejecta become more proton-rich 
because the electron neutrinos convert some fraction of neutrons 
to protons via the reactions $n + \nu_e \leftrightarrow p + e^-$.
Note again that $\nu_{e}$ capture is more efficient than $\bar{\nu}_{e}$ 
capture since the ejecta are neutron-rich.

\begin{figure}[t]
\epsfxsize=3.in
\leavevmode
\epsffile{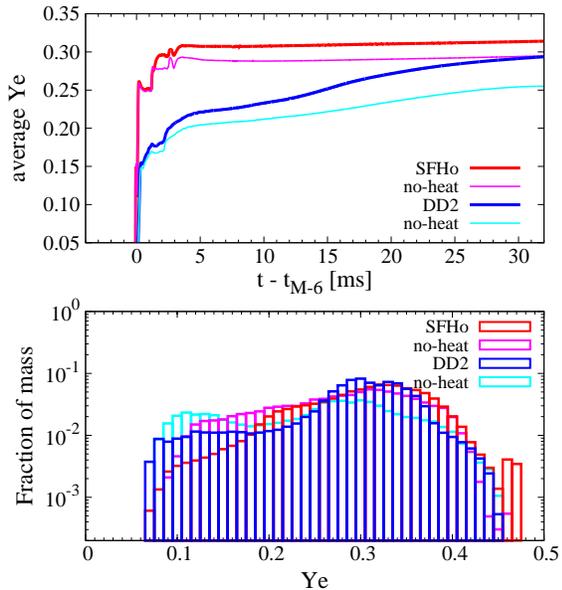}
\caption{ 
Same as Fig.~\ref{fig4} but for simulations with and without (denoted as no-heat) 
the neutrino heating for SFHo (red and magenta (no-heat)) and DD2 (blue and light blue (no-heat)).
\label{fig5}}
\end{figure}

Figure~\ref{fig5} compares the time evolution of $\langle Y_{e} \rangle$ (upper panel) 
and the mass-distribution histograms for $Y_e$ at $\approx 25$\,ms after 
the onset of merger (lower panel) between simulations with and without
neutrino heating for SFHo and DD2. 
The results indicate that for SFHo, $\langle Y_{e} \rangle$ is increased to be $\approx 0.29$ due to 
the positron capture and the neutrino heating pushes up it further by $\approx 0.02$ at the 
end of the simulations.
For DD2, the effect of the positron capture is weaker and the neutrino heating plays a relatively
important role, increasing $\langle Y_{e} \rangle$ by $\approx 0.03$.  
Such enhancements of $\langle Y_{e} \rangle$ due to the neutrino heating would be important 
in considering the $r$-process nucleosynthesis~\cite{Wanajo}.

The mass-distribution histograms also shift towards the higher $Y_{e}$ side 
due to the neutrino heating.
However, the distributions still show a broad feature even without the neutrino heating.
This suggests that the positron capture resulting from the strong shock heating
due to general relativistic gravity is primarily responsible for making the $Y_{e}$ 
distribution broad for DD2 and SFHo.
For much stiffer EOS like TM1, the neutrino heating would play a relatively major role.
Although our treatment for the neutrino transfer is an approximate one, our results
indicate that the neutrino heating plays an important role in determining the {\it chemical}
properties of the ejecta.

\section{Summary and discussion}

We have reported the first numerical
results of radiation hydrodynamics simulations in general relativity
focusing on the properties of the dynamical ejecta of the equal-mass BNS merger with typical mass
of each neutron star ($1.35M_\odot$).
Three modern finite-temperature EOS are employed to clarify the dependence 
of the ejecta properties on the EOS.  We found that the total mass of the ejecta 
is larger for softer EOS (giving smaller-radius neutron stars), and it exceeds
$0.01M_\odot$ only for the case that $R_{1.35} \alt 12$\,km, as
indicated in~\cite{Hotokezaka:2013iia}. 
As shown in~\cite{Li:1998bw,TH}, the electromagnetic luminosity of the ejecta 
by the radioactive decay of the $r$-process elements would depend sensitively 
on the ejecta mass, and hence, the predicted range of the luminosity spans in 
a wide range due to the uncertainty of the nuclear-matter EOS.

We also found that the averaged value of $Y_e$ of the ejecta is higher
for softer EOS like SFHo in which $R_{1.35}$ is smaller, reflecting the fact
that the shock heating is more efficient. 
For all of the models, the value of $Y_e$ for the ejecta has a broad distribution 
between $\sim 0.1$ and 0.45, by contrast with the previous studies~\cite{Rosswog,Bauswein}.
Here, both the strong shock associated with general relativistic gravity and 
the weak interactions play crucial roles for this.
Such a broad distribution may be well-suited for producing the universal~\cite{Sneden}
solar-abundance pattern of $r$-process elements as illustrated in~\cite{Wanajo}. 

For the EOS but for SFHo, the {\it dynamical} ejecta mass is of order $10^{-3} M_{\odot}$.
In this case, a rather higher merger rate of $\gtrsim 10^{-4}$ yr$^{-1}$
than the present estimates of the Galactic rate (a few $10^{-5}$ yr$^{-1}$)~\cite{Dominik}
is necessary to explain the amount of heavy $r$-process elements~\cite{Goriely,Qian}, 
if the the dynamical ejecta from binary neutron star mergers is responsible for their production.
In regards to this point, SFHo is an attractive EOS. We will study consequences of our 
results on the synthesis of heavy elements in the forthcoming paper.
If EOS is not very soft like SFHo, some other contributions, such as mergers of black hole-neutron star 
binaries~\cite{BHNS}, disk winds from accretion torus around a merger remnant 
black hole~\cite{Surman,Just2014}, 
and magnetorotational supernova explosions~\cite{Winteler} may be necessary.
In such cases, however, it is not clear whether the universality requirement can be
achieved or not.

In this work, we focused only on the equal-mass binary case and did not explore 
the dependence of the results on the binary parameters such as the total mass and the 
mass ratio. As reported in~\cite{Hotokezaka:2013iia}, the relative importance of the 
tidal interactions and the shock heating in the dynamical mass ejection depends on the 
binary parameters. It is interesting to explore the dependence of the results on binary 
parameters for SFHo and the resulting abundance profile in the future work, because
the observed abundance patterns of the metal-poor, $r$-rich stars show
some diversity in the lower mass-number region~\cite{Sneden}.
Also, we did not continue our simulations beyond $30$--40\,ms after the onset of merger.
For the longer time scales, magnetohydrodynamic processes~\cite{Kiuchi}, 
viscous heating, and nuclear recombination~\cite{Fernandez} could be important.
Self-consistent studies of these effects in the BNS merger also have to be done in the
future.

\section*{Acknowledgments}
We are grateful to M. Hempel for providing
the EOS table data and S. Wanajo for discussions. 
Numerical computations were performed on the
supercomputer K at AICS, XC30 at CfCA of NAOJ, FX10 at Information
Technology Center of Tokyo University, and SR16000 at YITP of Kyoto
University.  This work was supported by Grant-in-Aid for Scientific
Research (24244028, 24740163, 25103510, 25105508), for Scientific
Research on Innovative Area (24103001), by HPCI Strategic Program of
Japanese MEXT/JSPS (Project No. hpci130025, 140211).
Koutarou Kyutoku is supported by JSPS Postdoctoral Fellowship for
Research Abroad.



\end{document}